\def\ad#1#2#3{ad_#1^{(#2)}(#3)}

\def\square{\mathop{\mkern0.5\thinmuskip
                      \vbox{\hrule
                            \hbox{\vrule
                                  \hskip 6pt
                                  \vrule height 6pt width 0pt
                                  \vrule}%
                            \hrule}%
                      \mkern0.5\thinmuskip}}
\def\cH{{\cal H}} 
\def\const{{\rm const.\,\,}} 
\def\supp{{\rm supp}}
\def\ve{{\varepsilon}}
\def\vp{{\varphi}}
\def\vt{{\vartheta}}
\def\pt{{\partial_t}} 

\font\tit  =cmbx10  scaled\magstep1
\parindent=0pt
 
\centerline{\tit MINIMAL ESCAPE VELOCITIES}
\vskip .25in
\parindent=4cm
\item{W. Hunziker\ \ } Institut f\"ur Theoretische Physik, ETH Z\"urich
\item{}hunziker@itp.phys.ethz.ch
\medskip
\item{I.M. Sigal\ \ \ \ \ \ } Departement of Mathematics, University of Toronto
\item{}sigal@math.toronto.edu 
\medskip
\item{A. Soffer\ \ \ \ \ \ } Departement of Mathematics, Rutgers University
\item{}soffer@math.rutgers.edu  
\bigskip 
\centerline{(March 1997)}
\bigskip
\parindent=20pt 
\midinsert
\narrower
\noindent {\bf Abstract.} We give a new derivation of the minimal velocity estimates
[SiSo1] for unitary evolutions. Let $H$ and $A$ be self\-adjoint operators on a 
Hilbert space $\cH$. The starting point is Mourre's inequality $i[H,A]\ge\theta>0$, 
which is supposed to hold in form sense on the spectral subspace $\cH_\Delta$ of $H$ 
for some interval $\Delta\subset R$. The second assumption is that the multiple
commutators $\ad AkH$ are well-behaved for $k=1\ldots n\ \ (n\ge 2)$ . Then
we show that, for a dense set of $\psi$'s in $\cH_\Delta$ and all $m<n-1\,,$  
$\psi_t=exp(-iHt)$ is contained in the spectral subspace $A\ge\theta t$ as 
$t\to\infty$, up to an error of order $t^{-m}$ in norm. We apply this general result 
to the case where $H$ is a Schr\"odinger operator on $R^n$ and $A$ the dilation 
generator, proving that $\psi_t(x)$ is asymptotically supported in the set 
$|x|\ge t\sqrt{\theta}$ up to an error of order $t^{-m}$ in norm.

\endinsert
\parindent=0pt
\vskip .25in

\baselineskip=16pt

\noindent {\bf 1.\quad INTRODUCTION}
\medskip
Before posing the problem in abstract form we describe it in its
original concrete setting. Consider the Schr\"odinger equation
$$
i\pt\psi_t=H\psi_t;\qquad H={1\over 2}p^2+V(x)\quad{\rm on}\ \ L^2(R^n)
\eqno(1.1)
$$
for a particle in $R^n$ under the influence of a potential $V(x)$. We are
interested in the long--time behavior of orbits $t\to\psi_t$ in the
continuous spectral subspace $\cH_c$ of $H$. Under mild conditions on $V$, $H$
is selfadjoint and
$$
\langle p^2\rangle_{\psi}\le \const \langle H+c\rangle_{\psi}
\eqno(1.2)
$$
for some constant $c\in R$. Here $\langle\Phi\rangle_\psi=(\psi,\Phi\psi)$
denotes the expectation value of an observable $\Phi$ in the state $\psi$.
By Ruelle's theorem ([Rue], see also [CFKS], [HuSi1]) any orbit $\psi_t$ in 
$\cH_c$ is  {\it escaping} in a mean ergodic sense :
$$
\lim_{T\to\infty}\ {1\over T}\int_0^T dt\,\int_{|x|\le R}\ dx\ |\psi_t(x)|^2=0
\eqno(1.3)
$$
for any finite $R$. The question is : {\it how fast ?} The answer will of
course depend on the initial state $\psi$. The simplest example is the
free particle ($V=0$): if the Fourier transform $\widehat\psi$ of $\psi$
is smooth and supported outside some ball of radius $v>0$, then a standard 
asymptotic expansion gives the result
$$
\int_{|x|\le vt}\ dx\ |\psi_t(x)|^2\ =\ O(t^{-m})\quad (t\to\infty)
\eqno(1.4)
$$
for any $m$. In this sense the orbits $\psi_t$ of the given type are said to have a
{\it minimal escape velocity $v$} given by the support of $\widehat\psi$. 
To obtain a similar result for $V\neq 0$ we study the long--time behavior
of the expectation values $\langle A\rangle_t=(\psi_t\,,\,A\psi_t)$
for  suitable observables $A$, which evolve according to
$$
\pt\ \langle A\rangle_t\ = \langle i[H,A] \rangle_t\ .
\eqno(1.5)
$$ 
Mourre's very fruitful idea [Mou] was to find observables $A$ such that the commutator 
in (1.5) is {\it conditionally positive} , in the sense that
$$
E_\Delta\,i[H,A]\,E_\Delta\ \ge\ \theta\,E_\Delta\qquad (\theta>0)
\eqno(1.6)
$$
for some interval $\Delta\subset R$, where $E_\Delta$ is the corresponding
spectral projection of $H$. This implies that
$$
\langle\ A\ \rangle_t\ \ge\ \theta\,t\ +\ O(1)\ \to\ \infty\qquad (t\to\infty)
\eqno(1.7)
$$
for orbits $\psi_t$ in the spectral subspace $\cH_\Delta={\rm Ran}\,E_\Delta$.
Evidently $A$ must be unbounded, so that domain questions arise. Also
$\cH_\Delta$ must be a subspace of $\cH_c$ since $\langle A\rangle_t$
is constant and $\langle i[H,A]\rangle_\psi=0$ for any eigenvector
$\psi$ of $H$. (1.6) is a special case of a more general inequality due to
Mourre, which was first proven for  Schr\"odinger operators (including $N$--body 
systems), where $A$ was taken as the dilation generator :
$$
A\ =\ {1\over 2}\,(p\cdot x+x\cdot p)\ ;\quad 
i[H,A]\ =\ p^2-x\cdot\nabla V(x)
\eqno(1.8)
$$
([Mou], [PSS], see also [CFKS], [HuSi1]). In this case the intervals $\Delta$ 
for which (1.6) holds fill the continous spectrum of $H$, in the sense
that the corresponding subspaces $\cH_\Delta$ span  $\cH_c$.
Moreover, since  
$$
A=i\,[H,{1\over 2}x^2]
\eqno(1.9)
$$
is itself a commutator, (1.5) can be written as 
$\pt^2\langle x^2\rangle_t\ge 2\theta$
for orbits $\psi_t$ in $\cH_\Delta$, which implies that
$$
\langle x^2 \rangle_t\ \ge\ \theta\,t^2\ +\ O(t)
\quad (t\to\infty)\ .
\eqno(1.10)
$$
This is of course weaker than (1.4)\ : {\it it only says that the mean value of
$x^2$ for the probability distribution $|\psi_t(x)|^2$ diverges like 
$\theta\,t^2$, whereas we want to prove that the {\it support} of this distribution is
asymptotically contained in $|x|\ge t\,\sqrt{\theta}$ as $t\to\infty$} .
The first step is to derive a corresponding result for the spectral support
of $\psi_t$ with respect to $A$. We state this result in abstract form
for a pair $(H\,,\,A)$ of selfdjoint operators  on a Hilbert space $\cH$.
\medskip

\proclaim Theorem 1.1. 
Suppose that $\ad Ak{f(H)}$  is bounded for 
any $f\in C_0^\infty(R)$ and $k=1\dots n$, $n\ge 2$, and that the Mourre
inequality (1.6) holds for some open interval $\Delta\subset R$. Let
$\chi^\pm$ be the characteristic function of $R^\pm$. Then
$$
\|\chi^-(A-a-\vartheta t)e^{-iHt}g(H)\chi^+(A-a)\|\le \const t^{-m}
\eqno(1.11)
$$
for any $g\in C_0^\infty(\Delta)$, any $\vt$ in $0<\vt<\theta$ and any $m<n-1$, 
uniformly in $a\in R$\ .
\goodbreak
\bigskip

Since $g\in C_0^\infty (\Delta)$ and $a\in R$ are arbitrary, the vectors of the form
$$
\psi=g(H)\,\chi^+(A-a)\vp;\qquad \vp\in\cH
\eqno(1.12)
$$
are dense in $\cH_\Delta$. (1.11) says that, for any such $\psi$, 
$\psi_t=\exp(-iHt)\psi$ has spectral support in $[\,t\,\vt\,,\,+\infty)$
with respect to $A$ , up to a remainder of order $t^{-m}$ in norm.
\medskip\goodbreak
{\bf Remarks}
\medskip
{\it Commutators}. The hypothesis $\ad Ak{f(H)}\in L(H)$ may be replaced by
conditions on $\ad AkH$ which are more subtle to formulate since the operators
$A$ and $H$ are generally unbounded (see e.g. [ABG], [JMP]). For the
special case (1.8) this is further discussed below.
\medskip
{\it Resolvent smoothness and local decay}. We indicate briefly how minimal
velocity estimates are related to resolvent smoothness [JMP] and to local
decay [PSS]. Let $\rho(A)=(1+A^2)^{1/2}$. Setting $a=-\theta t/2$ and using that
$$
\rho(A)^{-\alpha}=\rho(A)^{-\alpha}\,\chi^\pm(A\pm ct)+O(t^{-\alpha})\ ,
$$
we obtain from (1.11)
$$
\|\rho(A)^{-\alpha}e^{-iHt}g(H)\rho(A)^{-\alpha}\|\le\const 
(1+t)^{-\min(\alpha,m)}\ . 
$$ 
For $\alpha,\ m>1$ this is integrable over $-\infty<t<+\infty$ , which (by Fourier
transform) leads to the resolvent estimate
$$
\sup_{z\notin R}\|\rho(A)^{-\alpha}(z-H)^{-1}g(H)\rho(A)^{-\alpha}\|\le\infty\ .
\eqno(1.13)
$$ 
Similar estimates for the derivatives with respect to $z$ (higher powers) of the
resolvent $(z-H)^{-1}$ are obtained using correspondingly higher values of
$\alpha, m$ (resolvent smoothness). Replacing $g$ by $g^2$ in (1.13) it also
follows that the operator $\rho(A)^{-\alpha}$ is $H$--smooth for $\alpha>1$
and therefore ([RSIV] Theorem XIII.25 and corollary)
$$
\int_{-\infty}^{+\infty}dt\,\|\rho(A)^{-\alpha}e^{-iHt}\psi\|^2\le \const
\|\psi\|^2\quad \forall \psi\in \cH_{\Delta'}\ ,
\eqno(1.14)
$$ 
where $\Delta'$ is any fixed compact subset of $\Delta$ (local decay). By an
independent argument of Mourre (given in [PSS]) the estimate (1.13) and 
therefore (1.14) can be improved from $\alpha>1$ to $\alpha>1/2$.
\bigskip
Our second result is an application of Theorem 1.1 to the Schr\"odinger
equation (1.1). Here $A$ is given by (1.8), and
$$
i^k\ad AkH = 2^{k-1}\,p^2+(-x\cdot\nabla)^k\,V(x) \ . 
\eqno(1.15)
$$
A simple way to satisfy the hypothesis of Theorem 1.1 in this case is to
assume that $V$ has relative bound less than 1 with respect to ${1\over 2}p^2$, 
and that the distributions $(x\cdot\nabla)^kV(x)$ are locally $L^2$ and (as 
multiplications operators) bounded relative to $p^2$ for $k=1\ldots n$. Then the 
operators (1.15) are bounded relative to $H$ and it is straightforward to compute and 
to estimate the the norms of $\ad Ak{(z-H)^{-1}}$ for ${\rm Im}(z)\neq 0$. 
Representing $f(H)$ by the resolvent $(z-H)^{-1}$ (e.g. using the Helffer-Sj\"ostrand 
formula ([HeSj], [Dav]) it then follows that $\ad Ak{f(H)}$ is bounded for $k=1\ldots 
n$. The result is that fast decay (large $m$) in (1.11) must be paid for by high 
smoothness of $V(x)$ for all $x$. This is unnatural, and in fact there is a better way
to construct $A$ which requires only smoothness of $V(x)$ for arbitray large 
$|x|$. The idea is to replace $x^2$ in (1.9) by a smooth, convex
function $G(x)$ which is equal to $x^2$ for large $|x|$ but constant
in some abitrary large ball $|x|\le R$. 
Then $A$ changes to 
$$ A = {1\over 2} (\nabla G(x) \cdot p + p\cdot \nabla  G(x) ) \ , $$ 
and the Mourre inequality can be established as before. Then the operators 
$\ad Ak{p^2} $ remain second order in $p$ with bounded coefficients, while                   
$$
i^k\,\ad AkV\ =\ (-\nabla G(x)\cdot\nabla)^kV(x)
$$
requires only derivatives of $V(x)$ in the region $|x|>R$. A more carful
construction of $G(x)$ due to Graf [Gra1] is specially adapted to the $N$-body case,
requiring only smoothness of the pair potentials for large separations (see [Skib], 
[Gri]). The following result can also be proven in this more general setting.
\medskip

\proclaim Theorem 1.2. 
Let $H$ and $A$ be given by (1.1) and (1.8). Then, under
the hypothesis of Theorem 1.1\ ,
$$
\| \chi^-(x^2- 2at - \vt\,t^2)\,e^{-iHt}g(H)\,\chi^+(A-a)\|\le \const t^{-m}
\eqno(1.16)
$$
for any $\vt$ in $0<\vt<\theta$, any $m<n-1$ and any $a\in R$.

\bigskip
We remark that for initial states of the form (1.13) this is equivalent to
$$
\int_{|x|\le v\,t} dx\,\|\psi_t(x)\|^2\ \le\ \const t^{-2m}
\eqno(1.17)
$$
for any $v<\sqrt\theta$. 
\bigskip
The proofs of these results are given in section 2. The main tool is the method 
of commutator expansions summarized in section 3.
\medskip
We conclude the introduction with some (not exhaustive) bibliographical notes.
Minimal velocity estimates  were first given by Sigal and Soffer [SiSo1] and then 
extended by Skibsted [Ski] and by G\'erard and Sigal
[GeSi], with applications to scattering theory ([SiSo2], [Sig], [HeSk]) 
and to the theory of resonances ([GeSi], [SoWe], [Nier]). Our derivation
is similar in spirit to [Ski] and incorporates remarks by Froese and Loss [FrLo].
The related subject of resolvent smoothness and local decay is more fully treated 
by time--independent methods e.g. in [PSS], [JMP] and [GIS], where further
references can be found. The generalization of Mourre's theorem using the
construction of Graf [Gra1] first appears in [Ski]. A simpler proof due to
Graf [Gra2] is given in [Gri]. For other applications of Mourre's inequality to 
wave equations and spectral geometry see e.g. [DHS], [DBiPr], [FHP]. 
\vglue 0.25in
\goodbreak
\noindent {\bf 2.\quad PROOFS}
\nobreak
\medskip

The following lemma gives the basic estimate for a proof by bootstrap of Theorem 1.1. 
A smooth function $f$ on $R$ is said to be of
order $p$ if for each $k$
$$
|f^{(k)}(x)|\ \le\ \const |x|^{p-k}\ .
$$ 

\proclaim Lemma 2.1. Suppose that $A,\ H$ and $g$ satisfy the hypothesis of
Theorem 1.1. Let $f$ be a positive $C^\infty$-function on $R$ of order $<4$ 
with $f' \le 0$ and $f(x) = 0$ for $x\ge 0$. Let $1\le s < \infty$; $a\in R$,  
$A_s = s^{-1} (A-a)$, and $\ve\le 1$. Then
$$
\eqalign{
g(H) i [ H , f(A_s)] g(H) 
&\le s^{-1} \theta g(H) f' (A_s) g(H)\cr
&+ s^{-(1+\ve)} g(H) f_1 (A_s) g(H) \cr 
&+ \const \  s^{-(2n-1-\ve)} g^2 (H)\cr } 
\eqno(2.1)
$$
uniformly in $a\in R$, where $f_1$ is a function on $R$ 
which again satisfies the hypothesis stated for $f$.
\par

\noindent{\bf Proof.} In the commutator $i[ H , f(A_s)]$ occurring in (2.1) we
can replace $H$ by a bounded operator
$$
H_b=H\,b(H)
\eqno(2.2)
$$
where $b\in C_0^\infty(R);\ b\equiv 1$ on $\supp(g)$. Then the commutators
$$ B_k = i \ \ad Ak{H_b} \ , \quad k=1\dots n\ ,$$ 
are bounded by hypothesis.  As a first step we show that
$$
i [ H_b , f(A_s) ]= -s^{-1} (-f' (A_s))^{1/2} B_1 (-f' (A_s))^{1/2} 
+\ {\rm remainder}\ .
\eqno(2.3) 
$$
Here and in the rest of the proof a {\it remainder} is defined as a quadratic form 
rem$(s)$ with an estimate 
$$ 
\pm {\rm rem}(s) \le s^{-(1+\ve)} f_1 (A_s) + \const \ s^{-(2n-1-\ve)} \ , 
\eqno(2.4)
$$
uniformly in $a$, where $f_1$ is a function on $R$
satisfying the hypothesis for $f$.  Any such remainder clearly
fits into (2.1) and needs no further consideration. 
To prove (2.3) we factorize $f= F^2$ and then expand the 
commutator 
$$ 
\eqalign{
i [H_b,f] &= i[H_b,F] F+ Fi [H_b,F] \cr 
&= \sum_{k=1}^{n-1} \ {1\over k!} 
s^{-k} (F^{(k)} B_k F + F B^*_k F^{(k)})\cr
&+ s^{-n} (R F + FR^*) \cr }
\eqno(2.5) 
$$ 
using (3.1).  Since $n\ge 2$ and since $F$ is of order $<2$ 
it follows from (3.2) that $R$ is bounded 
uniformly in $s$ and $a$.  Now we observe that all the terms 
in the expansion (2.5) except the leading term $(k=1)$ 
are remainders.  In particular
$$
|(\psi,F^{(k)}B_kF\psi)|\le \|B_k\|\,\|F^{(k)}\psi\|\,\|F\psi\|
\le \const\ (\psi,f_1\psi),
\eqno(2.6)
$$
where $f_1$ is a common upper bound  for $F^2=f$ and $(F^{(k)})^2$ which satisfies the 
hypothesis for $f$. The last term in (2.5) is estimated using the operator inequality
$$ \pm (P^* Q + Q^* P ) \le P^* P + Q^* Q $$
for $Q = s^{-{1\over 2} (1+ \ve)}F $; 
$P^* = s^{-n+{1\over 2} (1+\ve)} R $, 
with the result 
$$ \pm s^{-n} (RF + FR^*) 
\le s^{-(1+\ve)} f+ s^{-(2n-1-\ve)} \| R\|^2 \ . \eqno(2.7)$$
Therefore it remains to consider the leading term $(k=1)$ in (2.5), which 
is rewritten as
$$ s^{-1} (F' B_1 F+ F B_1 F') = -s^{-1} (v^2 B_1 u^2 + u^2 B_1 v^2) $$ 
by factorizing $F = u^2$, $-F' = v^2$. 
Since $u$ is of order $<1$ it follows  from (3.2) that $\| [B_1 ,u]\|
\le \const \ s^{-1}$ uniformly in $a$, and similarly 
for $[B_1,v]$.  As in (2.6) this leads to  the form estimate 
$$
\eqalign{ 
&s^{-1} |v^2 B_1 u^2 - uv B_1 uv|=s^{-1}|v[B_1,u]uv+v[v,B_1]u^2|\cr 
&\le\const \ s^{-2} (v^2 + u^2v^2+ u^4)\le \const \ s^{-2}\,f_1(A_s)\ ,\cr}  
$$
where $f_1$ shares the properties of $f=u^4$ (note that $v$ is of lower order than
$u$). Since the same estimate holds with $u$ and $v$ interchanged, we conclude that
$$
s^{-1}(F'B_1F+FB_1F')=-2s^{-1}\,uv\,B_1\,uv
$$
plus a remainder. (2.3) now follows since
$f' = 2FF' = -2 (uv)^2 $. 
\bigskip
To complete the proof of Lemma 2.1. we multiply (2.3) from both sides with
$g(H)=g(H)G(H)$, where $G\in C_0^\infty(\Delta)$ is real and
$G\equiv 1$ on $\supp(g)$. We also adjust the function $b$ in (2.2) such that 
$b\equiv 1$ on $\supp(G)$. Multiplying (1.6) by $G(H)$ from both sides we 
obtain the Mourre inequality\ :
$$
G(H)B_1G(H)=G(H)i[H,A]G(H)\ge \theta\,G^2(H).
\eqno(2.8)
$$
Abbreviating $G(H)=G$ and $(-f^\prime(A_s))^{1/2}=j$ we show that 
$$
s^{-1}GjBjG-s^{-1}jGBGj=
s^{-1}\bigl(\,jGB[j,G]+[G,j]BGj+[G,j]B[j,G]\,\bigr)
\eqno(2.9)
$$
is a remainder for any bounded $B=B^*$, using for $[G,j]$ the
expansion
$$
[G,j]=\sum_{k=1}^{n-1}{1\over{k!}}s^{-k}\,j^{(k)}(A)\,\ad AkG +s^{-n}R
$$
and its adjoint for $[j,G]$. Since $\ad AkG$ is bounded for 
$k\le n$, the right hand side of (2.9) then
becomes a sum of terms of the following types\ :
$$
\leqalignno{
&s^{-(k+l+1)}\left(j^{(k)}\,C\,j^{(l)}+j^{(l)}\,C^*\,j^{(k)}\right);
    \quad (0\le k,l\le n-1;\,k+l\ge 1);&(a)\cr
&s^{-(k+n+1)}\left(j^{(k)}\,C+C^*\,j^{(k)}\right);\hskip 1.3cm (0\le k\le n-1);&(b)\cr
&s^{-(2n+1)}\,C,&(c)\cr}
$$ 
where in each case $C$ stands for some operator which is bounded uniformly 
in $a$ and $s$. By the same arguments as in (2.6) and (2.7) these terms have 
corresponding upper and lower bounds
\goodbreak
$$
\leqalignno{
&\pm s^{-2}\,2\|C\|\,\left(j^{(k)2}+j^{(l)2}\right);&(a)\cr
&\pm \left(s^{-2}\,j^{(k)2}+s^{-2n}\,\|C\|^2\right);&(b)\cr
&\pm s^{-(2n+1)}\,\|C\|\ .&(c)\cr}
$$ 
For any bounded $B=B^*$ we therefore obtain 
$$
s^{-1}GjBjG\cong s^{-1}jGBGj\ ,
\eqno(2.10)
$$
meaning that the difference of the two expresssions is a remainder (2.4).
For $B=B_1=i[H_b,A]$ we use Mourre's inequality (2.8) to obtain from (2.3)\ :
$$
\eqalign{
G\,i[H,f]\,G\ &\cong\ -s^{-1} Gj\,B_1\,jG\ \cong\ -s^{-1}jG\,B_1\,Gj\cr
        &\le -s^{-1}\theta jG^2j\ \cong\ -s^{-1}\theta Gj^2G\cr 
        &=\ s^{-1}\theta Gf^\prime G\ ,\cr}
$$
with remainders arising 
from (2.3) and twice from (2.10). Multiplying from both sides with 
$g(H)$ removes $G(H)$ and leads directly to (2.1).
\hfill $\square$
\bigskip
\noindent {\bf Proof of Theorem 1.1.}
We prove a slightly stronger version of (1.11), which will serve 
later in the proof of Theorem 1.2. 
Let
$$ A_{ts} = s^{-1} (A-a - \theta t)\ ,$$
and suppose that $F$ is a positive $C^\infty$-function of order 
$\le 1/2$ on $R$ with $F' \le 0$ and $F(x) = 0$ for $x\ge 0$. 
Instead of (1.11) we show under the same hypothesis that 
$$ \| F(A_{ts}) e^{iHt} g(H) \chi^+ (A-a) \| \le \const  \ s^{-m}
\eqno(2.11) $$
for $m< n -1$, uniformly in $0\le t \le s$ and in $ a \in R$. 
(1.11) then follows by setting $t=s$ and by observing 
that, since $\vartheta<\theta$, 
$\chi^- (s^{-1} (A-a) - \vartheta) \le F (s^{-1} (A-a) - \theta) $
for some $F$ of the required type.  To prove (2.11) we consider the 
operator
$$ \phi_s (t) = g(H) f(A_{ts}) g(H) \ ; \ f= F^2 \ , $$ 
and the evolution 
$$ \psi_t = e^{-iHt} \chi^+ (A-a) \varphi \ ; \ \varphi \in {\cal H} \ . 
\eqno(2.12) $$
Then the estimate (2.11) to be proved reads 
$$
\langle \phi_s (t) \rangle_t = (\psi_t ,  \phi_s (t) \psi_t )
\le \const  \| \varphi \|^2 s^{-2m}\ ,  
\eqno(2.13) $$
uniformly in $1<s$ , $0\le t \le s$ and in $a\in R$. We compute 
$$ \eqalignno{
\partial_t \langle \phi_s (t) \rangle_t &= (\psi_t , D_t \phi_s (t) 
\psi_t ) \ ; &(2.14)\cr
D_t \phi_s (t) &= i [H,\phi_s (t) ] + \partial_t \phi_s (t) &\cr
&= g(H) i [ H,f(A_{ts}) ] g(H) - s^{-1} \theta g(H) f' (A_{ts}) g(H) \ .  &(2.15) \cr } $$
First we conclude that 
$$ \| D_t \phi_s (t) \| \le \const  \eqno(2.16) $$ 
uniformly in $s$, $t$, $a$, since $f$ is of order $\le 1$.
(cf. the remark after (3.4)).  Secondly, by (3.8), 
$$ \eqalign{
\langle \phi_s (0) \rangle_0 
&\le \| \varphi \|^2 \| F(s^{-1} (A-a) ) g(H) \chi^+ (A-a) \|^2 \cr
&\le \const  \ s^{-2n} \| \varphi \|^2\ . \cr }
\eqno(2.17) $$
Integrating (2.14) over $t$ and using (2.16) and (2.17) we find the  
crude estimate
$$ \langle \phi_s (t) \rangle_t \le \const  \| \varphi \|^2 (s^{-2n} + s ) $$
for $0\le t \le s$, which proves (2.13) for $m= -1/2$.  Now we bootstrap 
this estimate. First we note that by (2.15) and Lemma 2.1
$$
D_t \phi_s (t) 
\le s^{-1-\ve}  g(H) f_1 (A_{st}) g(H)\ +\ \const  s^{-(2n-1-\ve )} g^2 (H)\ .
\eqno(2.18) $$ 
As an induction assumption, suppose that (2.13) holds for some $m<n-1$. 
Since $f_1$ also satisfies the hypothesis for $f$ it then follows from (2.18) 
that 
$$ | \langle D_t \phi_s (t)\rangle_t | 
\le \const  \| \varphi \|^2 \cdot s^{-1}\bigl(s^{-(2m + \ve) } + 
s^{-(2(n-1)-\ve}\bigr)\ , $$ 
and again by integrating over $t$\ : 
$$ \langle \phi_s (t) \rangle_t \le \const  
\| \varphi \|^2\bigl(s^{-2n} + s^{-(2m+\ve)} + s^{-(2(n-1)-\ve)}\bigr) $$
uniformly in $0\le t\le s$ and in $a\in R$. Recalling that $\ve\le 1$, the best
decay for $s\to\infty$ is obtained by setting
$$
\ve\ =\ \min\,\bigl(1,\ (n-1)-m\bigr)\ ,
$$
which boosts the exponent $m$ in (2.13) to $m'= m+\ve/2$. Therefore (2.13) holds
for any $m<n-1$.\hfill $\square$
\bigskip
\noindent{\bf Proof of Theorem 1.2.}
It suffices to prove that 
$$ \| \chi (t^{-2} x^2 -2t^{-1} a - \vartheta ) e^{-iHt}
g(H) \chi^+ (A-a) \| \le \const  t^{-m}  \eqno(2.19) $$ 
if $\chi$ is a smoothed characteristic function 
of $(-\infty , -\ve ) $ for some $\ve > 0$ :
$\chi(x) = 1$ for large negative $x$, $\chi' \le 0$, 
and $\chi(x) = 0$ for $x\ge -\ve$.  Following 
the line of the previous proof we consider the operators
$$
\phi_s (t) = f (x^2_{ts}) ; \ f= \chi^2 ; \ 
x^2_{ts} = {1\over s^2} (x^2 - 2 at - \vartheta t^2) 
\eqno(2.20) $$
and the evolution 
$$ \psi_t = e^{-iHt} g(H) \chi^+ (A-a) \vp \ , \ 
\vp \in \cH \ , \eqno(2.21) $$
for $0\le t \le s$.  The desired
inequality (2.19) then reads
$$ \langle \phi_s (s) \rangle_s \le \const  s^{-2m } \ . \eqno(2.22) $$ 
Writing $x^2 - 2at - \vt^2 = (x- a)^2 - (a^2 + \vt) t^2$ we see that 
$$ \phi_s (t) = 0 \quad{\hbox{ for }}\quad t \le s 
\sqrt{ {\ve\over a^2 + \vt}} \equiv \alpha s \ , $$ 
and therefore
$$ \langle \phi_s (s) \rangle_s = 
\int\limits_{\alpha s}^s \ dt 
\langle D_t \phi_s (t) \rangle_s \ . \eqno(2.23) $$ 
To find $D_t \, \phi_s (t) $ we first compute 
$$ \eqalign{
i [ H, \phi_s (t) ] 
&= {i\over 2} [p^2 , \phi_s (t) ]
= s^{-2} \big(A f' (x^2_{ts} ) + f' (x^2_{ts}) A\big) \cr 
&= -2s^{-2} u(x^2_{ts}) A u(x^2_{ts}) \ , \cr } $$ 
where we have factorized $f' = -u^2$ and used that 
$\big[ [A,u] u\big] = 0$. 
Adding the term 
$$ \partial_t \phi_s (t) = 2s^{-2} (a+ \vt t) u^2 (x^2_{ts})$$
we arrive at 
$$ D_t \phi_s (t) = -2 t s^{-2} u(x_{ts}^{2}) (t^{-1} (A-a) - \vt) 
u(x^2_{ts}) \ . \eqno(2.24) $$ 
Now we use $\vt < \theta$ to estimate 
$$ \eqalign{
-(t^{-1} (A-a) - \vt) 
&\le -(t^{-1} (A-a)-\vt) \chi^- (t^{-1} (A-a) -\vt) \cr 
&\le\bigl(F(t^{-1} (A-a) - \theta )\bigr)^2 \cr } $$ 
by some smooth function $F$ of order $1/2$ supported in 
$R^-$ and with $F' \le 0$.  Setting 
$A_t = t^{-1} (A-a) - \theta $
we find
$$ \langle D_t \phi_s (t) \rangle_t \le 
2ts^{-2} \| F(A_t) u(x_{ts}^2 ) e^{iHt} g(H) \chi^+ (A-a) 
\vp \|^2 \ . \eqno(2.25) $$ 
On the other hand, it follows from (2.11) by setting 
$t=s$ that 
$$ \| F(A_t) e^{iHt} g(X) \chi^+ (A-a) \| \le \const  t^{-m} \ . 
\eqno(2.26) $$
Before we can use this estimate in (2.25) we must commute the factor 
$u(x^2_{ts})$ to the left.  The required commutator 
can be expanded to any order $n$\ : 
$$ [u(x^2_{ts}) , F(A_t) ] = \sum\limits_{k=1}^{n-1} 
{1\over k!} t^{-k} \ad Aku \ 
F^{(k)} (A_t)\ +\ t^{-n} R\ , $$
where $\| R \| \le \const  \| \ad Anu \| $. 
Since $u\in C^\infty_0 (R)$, the commutators $\ad Aku $ are 
easily bounded\ : 
$$ \eqalign{
-i \ad A1u 
&= x \cdot \nabla u(x^2_{ts}) = 2s^{-2} x^2 u' (x^2_{ts}) \cr 
&= 2x^2_{ts} u' (x^2_{ts}) + 2s^{-2} (2at + \vt t^2) u' (x^2_{ts}) \cr } $$ 
and so forth, with the result that 
$ \| \ad Aku\| \le \const $ uniformly in 
$1\le s<\infty$ and $0\le t \le s$.  Since (2.26) also holds if 
$F$ is replaced by a derivative $F^{(k)}$, we find the estimate
$$ \eqalign{
\langle D_t \phi_s (t) \rangle_s 
&\le \const  t s^{-2} (t^{-m} + \sum\limits_{k=1}^{n-1} 
t^{-k-m} + t^{-n})^2 \cr 
&\le \const  s^{-2} t^{-2m + 1} \cr } $$ 
for $1\le s < \infty$,  $0\le t \le s$, uniformly in $a$. 
Therefore, by (2.23), 
$$ \langle \phi_s (s) \rangle_s \le \const  s^{-2} 
\int\limits_{\alpha s}^s \ dt\  t^{-2m+1} \le \const  s^{-2m} \ . 
\eqno{\square}$$
\vglue .25in

\noindent {\bf 3.\quad COMMUTATOR EXPANSIONS}
\medskip

Let $H$ and $A$ be self-adjoint operators on a Hilbert space $\cH$ and suppose 
that $H$ is bounded.
To say that the commutor $i[H,A]$ is bounded means that the quadratic form
$$ i [ (H \psi , A\psi) - (A\psi , H\psi)] $$
on $D(A)$ is bounded and thus defines a bounded, symmetric operator called
$i[H,A]$.  In the same sense we assume that the higher commutators
$$ \ad AkH = [\ad A{k-1}H \ , \ A ] $$ 
are bounded for $k = 2 \dots n$. 
Let $f$ be a real $C^\infty$--function on $R$.  Then, under a further condition
given below, the commutator $[H,f(A) ]$ has an expansion 
$$ [H , f(A) ] = \sum\limits_{k=1}^{n-1} \ 
{1\over k!} f^{(k)} (A) \ad AkH\ +\ R_n 
\eqno(3.1) $$
with a remainder estimate
$$ \| R_n \| 
\le c_n \| \ad AnH \| 
\sum\limits_{k=0}^{n+2} \int dx (1+ |x|)^{k-n-1} 
| f^{(k)} (x) | \ .  
\eqno(3.2) $$
The further condition on $f$ is that the integrals (3.2) exist.  
The number $c_n$ is a numerical constant depending on $n$ 
but not on $f$, $A$ or $H$.  In particular, the expansion (3.1) holds if 
$$ f^{(k)} (x) = O(|x|^{n-\ve - k}) \qquad 
(x\to \pm \infty) \eqno(3.3)$$
for $k= 1\cdots n + 2$, i.e. if the function  $f(x) $ grows not faster 
than $|x|^{n-\ve}$, with corresponding slower growth of the successive 
derivatives.  We will refer to (3.3) by saying that $f$ is of order
$n-\ve$.  In that case (3.1) is defined in form sense on the domain of
$f^{(1)} (A)$. Taking the adjoint of (3.1) and noting that 
$$ \ad AkH^* = (-1)^k \ad AkH\ , $$ 
we also obtain
$$ [H,f(A)] = \sum\limits_{k=1}^{n-1} {1\over k!}\ . 
(-1)^{k-1} \ad AkH f^{(k)} (A)\ -\ R_n^* \ . 
\eqno(3.4) $$
This defines $[H,f(A)]$ as an operator on the domain of 
$f^{(1)} (A)= f' (A)$. In particular, if $f$ is of order $\le 1$ and $n\ge 2$,
then $[H, f(A)]$ is bounded.  In the general case where $H$ is not 
bounded, we will work with operators $g(H)$, $g\in C^\infty_0 (R)$, 
assuming that 
$$ \ad Ak{g(H)} \hbox{ is bounded for } k=1\cdots n \ . 
\eqno(3.5)  $$
Then, if $f$ is of order $<n$, 
$$ 
\eqalign{
[g(H) , f(A) ] 
&= \sum_{k=1}^{n-1} {1\over k!} 
f^{(k)} (A) \ad Ak{(g(H)}\ +\ R_n\ ;\cr 
\| R_n \| &\le \const  \| \ad An{(g(H)} \| \ ,\cr} 
\eqno(3.6)
$$ 
with a constant depending on $f$ and $n$, and similarly for the adjoint 
expansion.  All these formulas are particularly useful if the role of $A$ is
played by a scaled operator, say $s^{-1} A$, $0< s<\infty$.  Then the 
commutor expansions are expansions in powers of $s^{-1}$, e.g.
$$ [ g(H), f(s^{-1} A) ] 
= \sum\limits_{k=1}^{n-1} {1\over k!} s^{-k} 
f^{(k)} (A) \ad Ak{g(H)}\ +\ s^{-n} R_n \ . 
\eqno(3.7) $$ 
A simple but useful observation is the following.
Suppose that $f(x)= 0$ for $x\in R^+$, and let $\chi^+$ be the characteristic
function of $R^+$.  Then 
$$ \| \chi^+ (A) g(H) f(s^{-1} A) \| \le \const  s^{-n}\ . 
\eqno(3.8) $$

\noindent {\bf Proof.}
Since $\chi^+ (A) f (s^{-1} A) = 0$ we have 
$$ 
\chi^+ (A) g(H) f(s^{-1} A) = \chi^+ (A) [ g(H) , f(s^{-1} A) ] \ . $$ 
Inserting the expansion (3.7) we notice that only the remainder 
$s^{-n} R_n $ contributes, since 
$$
\chi^+ (A) f^{(k)} (A) = 0.\eqno{\square}
$$                                           
Commutator expansions of this type were introduced in [SiSo1] and have since become 
an important tool of operator  analysis. There are several versions 
([SiSo1], [Ski], [IvSi], [ABG]) which differ in the form of the remainder estimate. 
The results above  are derived in [HuSi2] and are based on the 
Helffer-Sj\"ostrand functional calculus ([HeSj], [Dav]). 
\bigskip

\noindent {\bf Acknowledgements.} 
This work was supported by the Swiss National Fund (WH), by 
NSERC under Grant NA 7901 (IMS), and by NSF (AS).  
\bigskip
\baselineskip=12pt
\centerline{\bf REFERENCES}
\vglue .25in
\parindent=30pt

\item{[ABG]}
W.O. Amrein, A. Boulet de Monvel and V. Georgescu\ :
$C_0$-Groups, Commutator Methods and Spectral Theory for $N$--Body
Hamiltonians. {\it Progress in Mathematical Physics, 
Vol.} {\bf 135}, Birkh\"auser Verlag (1996).
\smallskip

\item{[CFKS]}
H. Cycon, R. Froese, W. Kirsch and B. Simon\ : 
Schr\"odinger operators.
{\it Texts and Monographs in Physics}, Springer Verlag (1987).
\smallskip

\item{[Dav]}
E.B. Davies\ :
{\it Spectral Theory and Differential Operators.}
Cambridge University Press (1995).
\smallskip

\item{[DHS]} 
S. DeBi\`evre, P. Hislop and I.M. Sigal\ :
Scattering theory for the wave equation on non-compact manifolds.
{\it Rev. Math. Phys.} {\bf 4} (1992) 575--618.
\smallskip

\item{[DBiPr]}
S. DeBi\`evre and D.W. Pravica\ :
Spectral analysis for optical fibres and stratified fluids, I, II.
{\it J. Funct. Anal.} {\bf 98} (1991) and 
{\it Comm. P.D.E.} {\bf 17} (1992) 69--97.
\smallskip

\item{[FHP]}
R. Froese, P. Hislop and P. Perry\ :
A Mourre estimate and related bounds for hyperbolic manifolds with 
cusps of non-maximal rank.
{\it J. Funct. Anal.} {\bf 98} (1991) 292--310.
\smallskip

\item{[FrLo]}
R. Froese and M. Loss\ :
unpublished notes.
\smallskip

\item{[Ger]}
C. G\'erard\ : Sharp Propagation Estimates for N--Particle Systems.
{\it Duke Math. J.} {\bf 67} (1992) 483--515.
\smallskip

\item{[GIS]}
C. G\'erard, H. Isozaki and E. Skibsted\ : N-body resolvent estimates.
{\it J. Math. Soc. Japan} {\bf 48} (1996) 135--160.
\smallskip \  

\item{[GeSi]}
C. G\'erard and I.M. Sigal\ :
Space-time picture of semiclassical resonances.
{\it Comm. Math. Phys.} {\bf 145} (1992) 281--328.
\smallskip

\item{[Gra1]}
G.M. Graf\ : 
Asymptotic completeness for $N$-body short-range 
quantum systems\ : a new proof.
{\it Comm. Math. Phys.} {\bf 132} (1990), 73--101.
\smallskip

\item{[Gra2]}
G.M. Graf\ : private communication. 
\smallskip

\item{[Gri]}
M. Griesemer\ :
$N$--body quantum systems with singular 
interactions.
{\it Ann. Inst. H. Poincar\'e } (1998), to appear. 
\smallskip

\item{[HeSj]}
B. Helffer and J. Sj\"ostrand\ :
Equation de Schr\"odinger avec champ magn\'etique et \'equation de Harper.
In\ : Schr\"odinger operators.  
H. Holden, A. Jensen eds., 
{\it Lecture Notes in Physics Vol.} {\bf 345}
Springer Verlag (1989).
\smallskip

\item{[HeSk]}
I. Herbst and E. Skibsted\ :
Free channel Fourier transform in the long-range $N$-body problem.
{\it J. d'Analyse Math.} {\bf 65} (1995) 297--332.
\smallskip

\item{[HuSi1]}
W. Hunziker and I.M. Sigal\ :
The General Theory of N--Body Quantum Systems. In\ : Mathematical quantum theory\ :
II. Schr\"odinger operators, J. Feldman et al., eds.,
{\it CRM Proc. and Lecture Notes Vol.} {\bf 8}, Amer. Math. Soc. (1995).

\item{[HuSi2]}
W. Hunziker and I.M. Sigal\ : 
Time dependent scattering theory for $N$-body quantum systems.
Preprint, ETH Z\"urich (1997).
\smallskip

\item{[IvSi]}
V. Ivrii and I.M. Sigal,
Asymptotics of the ground state energies of large Coulomb systems,
{\it Annals of Math.} {\bf 138} (1993) 243--335.
\smallskip

\item{[JMP]}
A. Jensen, E. Mourre and P. Perry\ :
Multiple commutator estimates and resolvent smoothness in quantum scattering theory.
{\it Ann. Inst. H.Poincar\'e} {\bf 41} (1984) 207--225.

\item{[Mou]}
E. Mourre\ :
Absence of singular continuous spectrum for certain seladjoint operators.
{\it Commun. Math. Phys.} {\bf 78} (1981) 391--408.

\item{[Nie]}
F. Nier\ : 
The dynamics of some open quantum systems with short-range non-linearities.
Preprint, Ecole Polytechnique, Paris (1997).
\smallskip
\item{[PSS]}
P. Perry, I.M. Sigal and B. Simon\ : Spectral Analysis of $N$--body Shcr\"odinger
Operators, Ann. Math. {\bf 144}, (1981) 519--567.

\smallskip 
\item{[RSIV]}
M. Reed and B. Simon\ : 
Methods of Modern Mathematical Physics, IV: Analysis of Operators,
Academic Press (1978)
\smallskip  

\item{[Rue]}
D. Ruelle\ : 
A remark on bound states in potential scattering theory.
{\it Nuovo Cimento} {\bf A61} (1969) 655--662.
\smallskip

\item{[Sig]}
I.M. Sigal\ :  
On long range scattering.
{\it Duke Math. J.} {\bf 60} (1990) 473--496.
\smallskip

\item{[SiSo1]}
I.M. Sigal and A. Soffer\ : 
Local decay and velocity bounds. 
Preprint, Princeton University (1988).
\smallskip

\item{[SiSo2]}
I.M. Sigal and A. Soffer\ :
Long-range many-body scattering.
{\it Invent. Math.} {\bf 99} (1990) 115--143.
\smallskip

\item{[Ski]}
E. Skibsted\ : 
Propagation estimates for $N$--body Schr\"odinger operators.
{\it Comm. Math. Phys.} {\bf 142} (1992) 67--98.
\smallskip

\item{[SoWe]}
A. Soffer and M. Weinstein\ :
Time-dependent resonance theory. Preprint, Ann Arbor (1997).

\parindent=0pt
\end